\newcommand{\be}{\begin{eqnarray}}
\newcommand{\ee}{\end{eqnarray}}

\newcommand{\cn}{{\rm cn}}
\newcommand{\sn}{{\rm sn}}
\newcommand{\dn}{{\rm dn}}
\newcommand{\di}{\displaystyle}

\def\1{{\rm 1\mskip-4.5mu l} }

\def\R{\mathbb R}

\def\N{\mathbb N}
\def\Z{\mathbb Z}

\documentclass[12pt]{article}

\usepackage{amsmath,amssymb}
\usepackage{graphicx}
\usepackage{cite}
\usepackage{bm}
\begin{document}
\renewcommand{\thefootnote}{\fnsymbol{footnote}}

\vskip 15mm

\begin{center}

{\Large Supersymmetry vs  ghosts}

\vskip 4ex

D. \textsc{Robert}

\vspace{.5cm}

Laboratoire de mathematiques Jean Leray, UMR 6629, 
 Universit\'e de Nantes, 2 rue de la Houssini\`ere,
BP 92208, Nantes 44322, France

$\texttt{robert@math.univ-nantes.fr}$

\vspace{1cm} 

A.V. \textsc{Smilga}

\vspace{.5cm}

SUBATECH, Universit\'e de
Nantes,  4 rue Alfred Kastler, BP 20722, Nantes  44307, France
\footnote{On leave of absence from ITEP, Moscow, Russia.}

$\texttt{smilga@subatech.in2p3.fr}$
\end{center}

\vskip 5ex

\begin{abstract}
\noindent  We consider  the simplest nontrivial 
supersymmetric quantum mechanical system involving higher derivatives.  
We unravel the existence of additional bosonic and fermionic integrals of motion forming a nontrivial algebra.
This allows one to obtain 
the exact solution both in the classical and quantum cases. The supercharges $Q, \bar Q$ 
are not Hermitially conjugate to each other  anymore, which allows for the presence of negative energies
in the spectrum. We show that the spectrum 
of the Hamiltonian is unbounded from below. It is discrete and infinitely degenerate in the free oscillator-like
 case and becomes  continuous  running from $-\infty$ to $\infty$ 
when interactions are added. Notwithstanding the absence of the ground state, there is no collapse, which suggests that a unitary 
evolution operator may be defined.

\end{abstract}

\renewcommand{\thefootnote}{\arabic{footnote}}
\setcounter{footnote}0
\setcounter{page}{1}

\section{Introduction}
It was suggested in Refs.\cite{benign, TOE} 
 that the Theory of Everything may represent a 
conventional supersymmetric field theory involving higher derivatives and living in flat 
higher-dimensional space. Our Universe is associated then with a 3-brane classical solution 
in this theory (a kind of soap bubble embedded in the flat higher-dimensional
bulk), while gravity has the status of effective theory in the brane world-volume.

 Generically, higher-derivative theories involve ghosts \cite{PU} described usually as negative residues
of the propagator poles and/or indefinite metric of  Hilbert space. Speaking in more direct 
physical terms, the presence of ghosts means the absence of the lower bound (the ground state) 
in the spectrum of the Hamiltonian. This more often than not leads to violation of causality 
or unitarity or both (see e.g. the recent discussion in \cite{Dudas}).

 The problem of ghosts was discussed  recently in Refs.\cite{benign,dno,5d}. 
In particular, in Ref.\cite{dno}  a nontrivial quantum mechanical higher-derivative system was presented where
the spectrum was bounded from below and hence the ghosts were absent. To be more precise, the spectrum of this 
system has no bottom in the free ``Pais-Uhlenbeck oscillator'' case, but the bottom appears as soon as the interaction
(of a certain kind) is switched on. When the interaction constant $\alpha$ is small, the ground state energy
behaves as $ -C/\alpha$. Negative and large by absolute value, but finite.

This example was not supersymmetric, however, and the mechanism by which the ghosts were killed there seems 
to be specific for nonsupersymmetric systems. In Ref.\cite{5d}, we considered a supersymmetric model
($5D$ superconformal gauge theory reduced to 0+1 dimensions) which naively involves ghosts. But we showed
that one can effectively get rid of them, if working in  reduced  Hilbert space
where the Hamiltonian is Hermitian and its spectrum is bounded from below. It is supersymmetry which helps one to do it.
Indeed, the standard minimal supersymmetric algebra
 \be 
\label{algsusy} 
Q^2 = \bar Q^2 = 0, \nonumber \\
\{Q, \bar Q \} = 2H\ ,
 \ee
where the supercharges $Q, \bar Q$ are Hermitially conjugate to each other, 
implies that all eigenvalues of the Hamiltonian are non-negative and the ground state with zero or
positive energy exists. 

Though ghost-ridden, the model of Ref.\cite{5d} did not involve higher derivatives in the Lagrangian. 
The motivation
of the present study was to find out whether the ghost-killing mechanism found in \cite{5d} works also 
for higher-derivative supersymmetric theories.  To this end, we considered the  simplest
higher-derivative supersymmetric quantum mechanical system with the action
 \be
\label{supact}
S \ = \ \int dt d\bar \theta d\theta \left[ \frac i2 ( \bar {\cal D} X) \frac d{dt} ({\cal D} X) +
V(X) \right]
 \ee
($X$ is a real supervariable).
 We will be mainly interested in the case where $V(X)$ is a polynomial having the form 
 \be
 \label{V24}
 V(X) \ = \   - \frac {\omega^2 X^2}2 - \frac {\lambda X^4}4 \ .
 \ee
 We found that, though certain technical similarities between this system and the system considered in \cite{5d}
exist, the physics in this case is essentially different. In particular, there are no compelling
reasons to censor the negative energy states out of the spectrum. However, in spite of their presense
(so that the spectrum is unbounded both from above and from below), this does not lead to disaster.  
Irreversable loss of unitarity is usually related to collapse phenomenon where singularity is reached in finite time. 
In our case, there is no such collapse. Moreover, the eigenstates of the Hamiltonian that we found have real 
energies and it is reasonable to expect that the unitary evolution operator can be defined.

In the next section, we describe the model, write down the component expressions for the Lagrangian, supercharges and
the Hamiltonian. We  discuss the trivial noninteractive case $V(X) \propto X^2$ and then the generic case.
We exhibit the presense of certain additional integrals of motion, which makes the problem {\it exactly soluble}. 
In Sect. 3, we discuss the classical dynamics  and, for the  superpotential (\ref{V24}), write the solutions
to the classical equations of motion explicitly. We show that there is no collapse and the solution 
exists at all times. It has an oscillatory behavior with linearly rising
amplitude.  In Sect. 4, we address the quantum problem and find the   
spectrum and the eigenstates. In Sect. 5,
we consider a more complicated system where the higher-derivative term is added to the conventional kinetic term.
Its classical dynamics is even more benign than the dynamics of pure higher-derivative theory --- 
the amplitudes do not rise 
linearly anymore and the motion is bounded in a finite region of the phase space. 
The spectrum of the mixed system is probably discrete, but dense everywhere. In Sect. 6, we 
discuss briefly a model where still extra time derivative is added.   
The last section is devoted as usual to concluding remarks and speculations.

\section{The model.}

Let us express the action (\ref{supact}) in components. To this end, we substitute there
$$ X = x + \theta \bar \psi + \psi \bar \theta + D \theta \bar \theta \ ,$$
 $$ 
{\cal D} = \frac \partial {\partial \theta} + i\bar \theta \frac \partial {\partial t}, \ \ \ 
\bar  {\cal D} = -\frac \partial {\partial \bar \theta} - i\theta \frac \partial {\partial t}\ ,
 $$
and integrate over $d\bar \theta d\theta$. We obtain
 \be
\label{LHD}
L = \dot x \dot D + V'(x) D + V''(x)\bar \psi \psi + 
\dot {\bar \psi} \dot \psi \ .
 \ee
Note that this Lagrangian involves twice as much physical degrees of freedom compared to  standard
Witten's supersymmetric quantum mechanics \cite{WitSQM},
 \be
\label{Lstand}
L_{\rm stand} \ =\ \int d\bar \theta d\theta \left[ \frac 12 \bar  {\cal D} X   {\cal D} X + V(X)\right] =  \nonumber \\
\frac {\dot x^2 + D^2}2 + \frac i2\left(\dot \psi \bar \psi - \psi \dot{\bar \psi} \right) + V'(x)D + V''(x)\bar \psi \psi \ .
 \ee
Indeed, the field $D$ enters the Lagrangian (\ref{LHD}) with a derivative and becomes dynamical. In addition, $\bar \psi$ 
does not coincide anymore with 
the canonical momentum of the variable $\psi$, but represents a completely independent complex fermion
variable not necessarily conjugate to $\psi$. It is convenient to denote it by 
$\chi$ and reserve the notation
$\bar \psi, \bar \chi$ for the canonical momenta
$$\bar \chi \equiv ip_\chi = i\dot \psi,\ \ \ \ \ \ \bar \psi \equiv ip_\psi = -i\dot \chi \ .$$
 Introducing also
$$p \equiv p_x = \dot D; \ \ \ \ \ \ \ \ P \equiv p_D = \dot x\ ,$$
we can derive the canonical Hamiltonian
\be
\label{H}
H =pP -DV'(x) + \bar \psi \bar \chi - V''(x) \chi \psi\ .
 \ee
The Lagrangian (\ref{LHD}) (with $\chi$ substituted for $\bar \psi$) 
 is invariant (up to a total derivative) with respect to 
the supersymmetry transformations,
\be
\label{susytran}
\delta_\epsilon x \ =\ \epsilon \chi + \psi \bar \epsilon\ , \nonumber \\
 \delta_\epsilon \psi \ =\ \epsilon (D - i\dot x)\ ,  \nonumber \\
\delta_{\bar \epsilon} \chi \ =\ \bar\epsilon (D + i\dot x)\ , \nonumber \\
\delta_\epsilon D \ =\ i(\epsilon \dot \chi - \dot \psi \bar \epsilon)\ .
 \ee
The corresponding N\"other supercharges are 
 \be
\label{Qcan}
Q \ =\ \psi[p + iV'(x) ] - \bar \chi (P - iD)\ , \nonumber \\
\bar Q \ =\ \bar \psi (P+iD) - \chi [p - iV'(x)]\ .
 \ee
One can be convinced that the algebra (\ref{algsusy}) holds, but, in contrast
to the standard SQM, $Q$ and $\bar Q$ are not Hermitially conjugate to each other. This
is the main reason for all the following complications.

Consider the simplest case,
 \be
\label{Vomx2}
V(X) = - \frac {\omega^2 X^2}2 \ .
 \ee
It is convenient to make a canonical transformation   
 \be
\label{zamena}
 x = \frac {x_+ + x_-}{\sqrt{2\omega}}, \ \  D =  \sqrt{\omega/2} (x_+ - x_-), \ \ p =  \sqrt{\omega/2}(p_+ + p_-), \ \ 
 P = \frac {p_+ - p_-}{\sqrt{2\omega}}\ , \nonumber \\
 \psi = \frac {\psi_+ + \psi_-}{\sqrt{2\omega}}, \ \   \chi = \frac {\bar\psi_- - \bar\psi_+}{\sqrt{2\omega}}, \ \ 
 \bar \psi  = \sqrt{\omega/2} (\bar\psi_+ + \bar\psi_-),\ \ 
 \bar \chi  = \sqrt{\omega/2} (\psi_- - \psi_+) \ .
 \ee
In terms of the new variables $x_\pm, p_\pm, \psi_\pm, \bar\psi_\pm$, the supercharges and Hamiltonian
acquire a simple transparent form
  \be
\label{Qpm}
Q \ =\ \psi_+ (p_+ - i\omega x_+) + \psi_- (p_- - i\omega x_-) \equiv Q_+ + Q_- \ , \nonumber \\
\bar Q \ =\ \bar \psi_+ (p_+ + i\omega x_+) -  \bar\psi_- (p_- + i\omega x_-) \equiv \bar Q_+  -  \bar Q_-\ ;
  \ee
\be
\label{Hpm}
 H = \frac {p_+^2 + \omega^2 x_+^2}2 + \omega \psi_+ \bar \psi_+ - 
 \frac {p_-^2 + \omega^2 x_-^2}2 - \omega \psi_- \bar \psi_-  \equiv H_+ - H_- \ .
 \ee

In other words, the system represents a combination of two independent supersymmetric oscillators
such that the energies of the second oscillator are counted with the negative sign (In nonsupersymmetric case, such
a system was first considered by Pauli back in 1943 \cite{Pauli}).
 The states are characterized by quantum
numbers $\{n_\pm, F_\pm \}$, where $n_\pm$ are nonnegative energies characterizing the excitation levels 
of each oscillator and $F_\pm = 0,1$ are the fermion numbers, the eigenvalues of the operators $\psi_\pm \bar \psi_\pm$.
The spectrum of the Hamiltonian
 \be
\label{specosc}
E_{n_+, n_-} =\ \omega(n_+ - n_-)
 \ee
is infinitely degenerate at each level depending neither on $n_+ + n_-$ nor on $F_\pm$. The spectrum (\ref{specosc}) is discrete
involving both positive and negative energies. 

We see that, in spite of supersymmetry, the spectrum has no bottom and hence involves ghosts. In contrast to
what was the case for $5D$ superconformal theories \cite{5d}, 
the negative energy states have the same multiplet structure
as the positive energy ones and there are no ``scientific'' reasons 
(i.e. the reasons based on certain symmetry considerations) to exclude these states from the spectrum.

  However, these ghosts are definitely of {\it benign} variety. Actually, when the system
consists of several noninteracting subsystems whose energies are individually conserved, the sign with 
which these energies are counted
in the total energy is a pure convention. The problems may (and do usually) arise when the subsystems 
start to interact. 
Then, if it is the difference rather than the sum of the energies of individual subsystems that is conserved, 
there is a risk that the 
individual energies would rise indefinitely leading to the collapse with associated unitarity and causality loss.

What happens in our case ?
A proper way to include interactions is to modify the superpotential (\ref{Vomx2}). The key observation is that for any superpotential $V(X)$
the system involves besides $H, Q, \bar Q$ two extra even and two extra 
odd conserved charges. They can be chosen 
in the form 
 \be
\label{NFT}
 N \ = \ \frac {P^2}2 - V(x) \ , \nonumber \\
F \ =\ \psi \bar \psi - \chi \bar \chi\ , \nonumber \\
 T \ = \ \psi[p - iV'(x)] + \bar \chi(P+iD) \ , \nonumber \\
 \bar T \ =\ \bar \psi(P-iD) + \chi [p + iV'(x)]\ .
  \ee
The superalgebra $(H,N,F ; Q, \bar Q, T, \bar T)$ has the following nonvanishing
commutators:
 \be
\label{supalgebra}
\{Q, \bar Q\} = \{T, \bar T \} = 2H; \nonumber \\
 \ [\bar Q, F] = \bar Q,\  [Q, F] = -Q,\ [T, F] = -T,\ [\bar T, F] = \bar T;
\nonumber \\
\ [Q, N] = [T, N] = \frac {Q-T}2, \ \ -[\bar Q, N] = [\bar T, N] = \frac{\bar Q 
+ \bar T}2 \ .   
 \ee
Now, $T$ and $\bar T$ are the extra supercharges, 
the subalgebra involving the operators
$(H; Q,\bar Q, T, \bar T)$ coincides with the standard subalgebra of extended ${\cal N} =2$
supersymmetry ${\cal S}_2$.
\footnote{${\cal S}_2$ is an ideal of the superalgebra (\ref{supalgebra}) and hence the 
latter is not simple. It represents a semidirect sum of the Abelian Lie algebra $(F,N)$ and
${\cal S}_2$. }
 This leads to 4-fold degeneracy of each nonvacuum level in quantum 
problem (but does not lead necessarily to positivity of their energies as $Q$ 
is not conjugate to $\bar Q$ and $T$ is not conjugate to $\bar T$). 
$F$ is the operator of fermion charge. As defined, it takes values $0$ for the states with
the wave functions 
$\Psi \propto 1$ and $\Psi \propto \psi\chi$, the value $1$ for the states $\Psi \propto \psi$ and the value $-1$ for the states
$\Psi \propto \chi$. (The convention is somewhat unusual, 
but one could bring it to the standard form by interchanging 
$\chi$ and $\bar \chi$.)   Finally, 
the conserved charge $N$ is a new object that is specific for the problem in hand. 
The corresponding symmetry   of the  action is
 \be
 \label{symact}
D \to D + \alpha \dot x\ .
  \ee
Indeed, the Lagrangian (\ref{LHD}) is shifted by a total derivative after this transformation. 

\section{Classical dynamics.}
Let us disregard the fermion variables and concentrate on  the dynamics of the 
bosonic Hamiltonian
 \be
\label{Hbos}
H_B \ =\ pP - DV'(x)\ .
 \ee
It involves two pairs of canonic variables. The presence of the extra integral of 
motion $N$ implies that the system is exactly soluble and seems to imply that the 
variables can be separated and the classical trajectories represent toric orbits.
This would be true for a conventional system with positive definite kinetic term. In our case, the situation
is more complicated. To begin with, the variables cannot be easily separated.  
Indeed, excluding the momenta from the corresponding canonical equations of motion, 
we obtain
 \be
 \label{eqmot}
 \ddot x -V'(x) = 0; \ \ \ \ \ \ \ \ \ \ddot D - V''(x) D = 0 \ .
 \ee
 The equation for $x$ does not depend on $D$, but the equation for $D$ {\it does}
 depend on $x$ for generic $V(x)$.

Let us try first to add the cubic term to the superpotential $V(X)$. As we see, 
the same function taken with the negative sign plays the role of the potential 
for the variable $x$. If $V(x) \propto x^3$ at large $x$, the potential is not binding
and the motion is infinite such that infinity is reached at a finite time. This 
is the collapse signalizing the presence of the ghost of malignant variety. 

Let us choose now the polynomial negative definite superpotential. It involves only the even powers of $X$. 
The simplest nontrivial
case is  Eq.(\ref{V24}).  
The potential is confining now and the equation of motion has a simple solution
representing an elliptic cosine function with the parameters depending on the integral
of motion $N$, 
 \be
 \label{solforx}
 x(t) \ =\ 
x_0 \, \cn [\Omega t , k]
 \ee
with 
  \be
 \label{param}
\alpha = \frac {\omega^4}{\lambda N},\ \ \Omega = [\lambda N (4+\alpha)]^{1/4},\ \ \ 
k^2 \equiv  m = \frac 12 \left[ 1 - \sqrt{\frac \alpha{4+\alpha}} \right], \nonumber \\
x_0 = \left( \frac {N}{\lambda} \right)^{1/4} \sqrt {\sqrt{4+\alpha} - 
\sqrt{\alpha}}
 \ .
  \ee
Here $k$ is the parameter of the Jacobi elliptic functions. \cite{ellipt} 
 \footnote{ Recall that,  if $k \in ]0, 1[$ and $t=\int_0^\phi\frac{d\theta}
{\sqrt{1-k^2\sin^2\theta}}$, then the elliptic functions are: $\sn\, t = \sin\phi$, 
$\cn \, t =\cos\phi$, $\dn \, t = \sqrt{1-k^2\sin^2\phi}$.
   The functions $\sn, \cn, \dn$ are periodic with period $4K$ where 
$K= \int_0^{\pi/2}\frac{d\theta}{\sqrt{1-k^2\sin^2\theta}}$.}  

The equation for $D$ represents an elliptic variety of the Mathieu equation. Generically, it is not
the simplest kind of equations, but in our case the solutions can be obtained in a rather explicit form.
\footnote{We thank  N. Nekrasov for this remark.}
One of the solutions is
 \be
\label{D1t}
D_1(t) \propto \dot x(t) =  \propto \sn \left[\Omega t, k\right] \, \dn 
\left[\Omega t, k \right]\ .
 \ee 
The second solution can be found from the condition that the time derivative of 
the Wronskian $W = D_1 \dot D_2 - D_2 \dot D_1$ vanishes. We find 
    \be
\label{D2t}
D_2(t) \propto \dot x(t) \int^t \frac {dt'}{\dot x^2(t')}  =  \propto  
\sn \left[\Omega t, k\right] \, \dn \left[\Omega t, k \right]\,
 \int^t \frac {dt'}{\sn^2 \left[\Omega t', k\right] \, \dn^2 \left[\Omega t', k \right]} \ .
 \ee 
When $\omega = 0$, the integral in Eq.(\ref{D2t}) can be done analytically and we obtain  
 \be
 \label{solD}
D(t) = A\, \sn \left[\Omega t, \sqrt{1/2}\right] \, \dn 
\left[\Omega t, \sqrt{ 1/2}\right] + \nonumber \\
B \left\{ \cn\left[\Omega t , \sqrt{1/2}\right] - 
\Omega t\, 
 \sn \left[\Omega t, \sqrt{1/2}\right] \, 
\dn \left[\Omega t , \sqrt{1/2} \right]\, \right\}
 \ee
Two independent solutions (\ref{D1t}), (\ref{D2t}) exhibit oscillatory behaviour with constant and 
linearly rising amplitude, correspondingly (see Fig.1).    
 \footnote{
One can remind the situation for the ordinary Mathieu equation. In generic case, its
solutions, the Mathieu functions, exhibit oscillatory behaviour with the amplitude that either
 oscillates itself or rises exponentially. But for some special characteristic values of parameters, 
the amplitude stays   constant or rises
lineary, like in our case.} The energy $E$ does not depend on $A$ and is
 \be
\label{EBN}
E = B\lambda^{1/4}(4N)^{3/4}\ .
 \ee

\vspace{-3cm}

\begin{figure}[h]
   \begin{center}
 \includegraphics[width=7.5in]{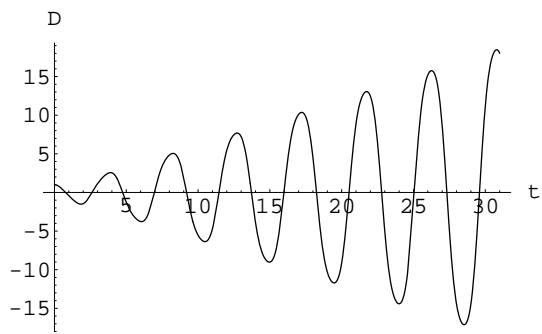}
        \vspace{-18cm}
    \end{center}
\caption{The solution of the equation (\ref{eqmot}) for $D(t)$ with the parameters
 $\omega = \lambda = N = 1$
and inital conditions $D(0)=1, D'(0)=0$.}
\label{Dlinrost}
\end{figure}

%\vspace{20cm}

%\newpage

\section{Quantum dynamics}

\subsection{Bosonic system}

Consider first the bosonic Hamiltonian (\ref{Hbos}). Let us prove that the corresponding evolution operator is
unitary. To this end, 
it is convenient to  perform a partial Fourier transform and consider the wave function in
the mixed representation,
  \be
 \label{Fourrier}
\tilde \Psi(x,P)\ = \ (2\pi)^{-1/2}\int_{-\infty}^\infty {\rm e}^{-iPD}\, \Psi(x,D) \, dD,
 \ee
The Schr\"odinger equation 
$$
i \frac {\partial  \Psi}{ \partial t} \ =\ K_B  \Psi
$$
 ($K_B$ is the operator obtained from $H_B$ by the corresponding canonical transformation)
for the function $\Psi(x,P)$ (we will not right tildas anymore) represents a linear first order
 differential equation,
  \be
\label{Sch1}
  \frac{\partial \Psi}{\partial t} + P\frac{\partial \Psi}{\partial x} + 
V^\prime(x)\frac{\partial  \Psi}{\partial P} = 0
  \ee
This equation can be easily solved by the charateristics method \footnote{see for example Ref.\cite{CourHil}}. 
The characteristic system is here
\be\label{caract}
\dot x &=& P \nonumber \\
\dot P &=&   V^\prime(x)
\ee
The equations (\ref{caract}) represent
 a half of original Hamilton equations of motion for the system (\ref{Hbos}). They can be interpreted as the
  Hamilton equations for the system described by the ``Hamiltonian''  
$ {P^2}/{2} - V(x)$. The latter  coincides with the extra integral of motion $N(P,x)$ defined before and should not 
be confused   with the true Hamiltonian $H_B$.

Let us denote by $\Gamma^t$ the flow determined by  (\ref{caract}).
 By definition, we have  $\Gamma^t(x_0, P_0) = (x_t, P_t)$.  We  clearly see that the Schr\"odinger equation
(\ref{Sch1}) is solved by 
\be\label{Sch2}
\  \Psi_t(x,P) = \  \Psi_0(\Gamma^{-t}(x,P))
\ee
 with an arbitrary  $\  \Psi_0(x,P)$. 
 Moreover, as $-V$ is confining, the flow 
  $\Gamma^t$ is well defined everywhere in $\R^2$  for all  times  and this property  entails that the Hamiltonian $K_B$ and
 hence   $H_B$  are  essentially
   self-adjoint.
\footnote{this  means that $K_B$, $H_B$ have a unique self-adjoint  continuation  in $L^2(\R^2)$ starting from the space of 
smooth, finite support functions on $\R^2$.}

At the next step, we will solve the stationary spectral problem for $H_B$ and find the eigenstates. We will construct
the states where not only the Hamiltonian $H_B$, but also the operator $N$ have definite eigenvalues.
\footnote{A note for purists: as most of these states belong to continuum spectrum, they 
represent {\it generalized} eigenstates of $H_B$ and $N$.}
 The system is integrable
and a regular way to solve it is to go over into action-angle variables. There is some specifics in our case.
We will follow the standard procedure not for $H_B$ (it is not possible as the variables cannot be separated there)
 but for the quasi-Hamiltonian $N$ involving only one pair of variables $(P,x)$. 
  Thus, we perform  a canonical transformation
   $S$: $(x,P)\mapsto (I, \varphi)$,  ($I$  is the action variable, $\varphi$ is the angle, $I \in \, ]0, +\infty[$,
   $\varphi\in \, [0, 2\pi[$)  such that in this new coordinates system the flow is 
   \be
   \Gamma^t(S^{-1}(I,\varphi)) = S^{-1}(I, \varphi+t\sigma(I))
     \ee
  where $\sigma(I) =  \partial N/ \partial I$.    Let us recall that $I$ is  given by the following integral
  $$I = \frac 1{2\pi} \oint P dx = \ \frac 1{2\pi} \int_{N(x,P)\leq N_0}dxdP\ , $$ 
where $N_0$ is the energy coinciding in our case with the value of the integral $N$ on the trajectory.  
For the potential (\ref{V24}), one can derive
 \be
\label{sigma}
 \sigma \ =\ \frac {\pi \Omega}{2K(k)}\ ,
  \ee
with $\Omega,k$ written in Eq.(\ref{param}). 
In the purely quartic case, $\omega = 0, \lambda = 1$,
 \be
\sigma =  \left( \frac {3I \pi^4}{16 K^4} \right)^{1/3} \ =\ \frac {\pi  N^{1/4}}{\sqrt{2} K}
 \ee
with
$$K \equiv K(1/\sqrt{2})\ =\ \frac {\Gamma^2(1/4)}{4 \sqrt{\pi}} \approx 1.85 \ .$$
   
The explicit expressions for the canonical transformation $S$ from  the action-angle variables
to the variables $x,P$   are in  this case
   \be
\label{cantran}
   x  &=& \Omega (I) \, \cn\left(\frac{2K}{\pi}\varphi\right) \nonumber\\
   P &=& -\Omega^2(I) \, \sn\left(\frac{2K}{\pi}\varphi\right)\dn\left(\frac{2K}{\pi}\varphi\right).
   \ee
 with the angle  $\varphi\in \, \R/2\pi\Z$ and the positive  action $I>0$.

   In the representation where the wave function $\Psi$ depends on $I$ and $\varphi$, the solution (\ref{Sch2}) 
to the Schr\"odinger equation takes the form 
  $$\Psi_t(I,\varphi) \equiv U(t) \Psi_0(I,\varphi) =  \Psi_0 (I, \varphi- t\sigma(I))\,.$$
 In this representation, 
$U(t)$ is a unitary evolution in the
  Hilbert space $L^2(]0,\infty[, \R/2\pi\Z)$. Its generator is a new quantum Hamiltonian:
 \be
\label{HamIphi}  
{\cal H}\psi = -i\sigma(I)\frac{\partial\Psi}{\partial\varphi}\, .
 \ee
The  Hamiltonians $H_B$ and $K_B$ are unitary  equivalent to the Hamiltonian
  ${\cal H}$. 

\begin{figure}[h]
   \begin{center}
 \includegraphics[width=9.0in]{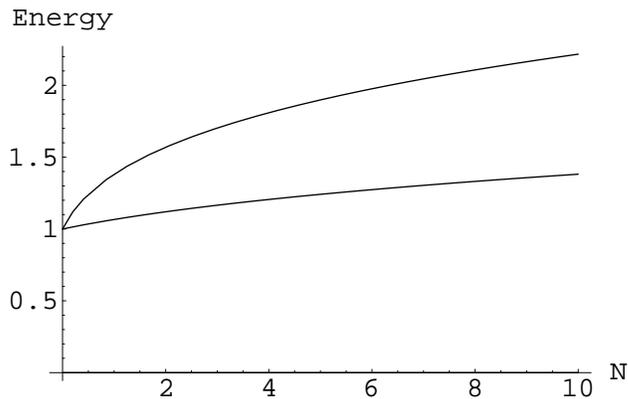}
      \vspace{-22cm}
  \end{center}
  \caption{The dependence $E_1(N)$. The lower curve corresponds to the choice $\omega = 1, \lambda = 0.1$ and the upper one
  to $\omega = \lambda = 1$.}
  \label{EotN}
  \end{figure}

Using a Fourier decomposition in the variable $\varphi$,  we have an explicit spectral decomposition for ${\cal H}$.
  If $\di{\Psi(I,\varphi) = \sum_{n\in\,\Z}\Psi_n(I){\rm e}^{in\varphi}}$, then
  \be\label{specrep}
  {\cal H}\psi(I,\varphi) = \sum_{n\in\,\Z}n \sigma(I)\Psi_n(I){\rm e}^{in\varphi}.
  \ee
Substituting in $E_n = n\sigma(I)$ the expression (\ref{sigma}), we derive the quantization condition 
  \be
\label{kvanty}
E_n\ = \ \frac {\pi n}{2 K(k)} \left[ \lambda N \left(4 +
\frac {\omega^4}{\lambda N} \right) \right]^{1/4}\ .
 \ee
  In the limit $\lambda \to 0$, the dependence of 
the left hand side
of Eq.(\ref{kvanty}) on $N$ disappears and we reproduce the simple oscillator quantization condition
$E = \omega n$ coinciding with Eq.(\ref{specosc}). When $\lambda \neq 0$, the right hand side of Eq.(\ref{kvanty})
depends on $N$ [$E_n \sim \pi n (\lambda N)^{1/4}/[\sqrt{2} K(1/\sqrt{2})]$ for large $N$] and 
only a certain combination of $E$ and $N$ is quantized, but not the energy by itself.
  For illustration, the function $E_1(N)$ is plotted
in Fig.\ref{EotN} for two choices of parameters.
The dependence of $\sigma$ on $I$ and hence $E_n$ on $N$ reveals that the spectrum is continuous here, with eigenvalues lying 
in two intervals  $]-\infty, -\omega]\cup[\omega, +\infty[$  plus  the eigenvalue $\{0\}$. The same qualitative picture (continuum spectrum 
which can be supplemented by isolated eigenvalues) 
holds for generic binding potentials $-V(x)$, in particular, for generalized anharmonic oscillators, 
 $V(x) = -a_0x^{2\ell} + a_1x^{2\ell-1} + \cdots +a_{2\ell}$, $a_0>0$,   $\ell >1$, where  
$\sigma(I) \sim I^{\frac{\ell-1}{\ell+1}}$ for large $I$. 

The generalized eigenfunctions of the Hamiltonian (\ref{HamIphi}) are labelled by the 
parameters $I_0 \in \, \R$  and  $n \in \, \Z$,
 \be
\label{PsiIphi} 
\Psi_{I_0n} (I,\varphi) \ = \ \delta(I-I_0) e^{in\varphi}\ .
 \ee
Going back to the original variables using Eqs.(\ref{Fourrier},\ref{cantran}), we obtain
 \be
\label{PsiEN}
\Psi_{EN}(x,D) \ =\ \frac 1{\sqrt{N + V(x)}} e^{iS(x,D)}\ ,
 \ee
where
\be
\label{solHamJac}
S(x,D) = D\sqrt{2[N + V(x)]} + \frac E {\sqrt{2}} \int^x \frac {dy}{\sqrt{N + V(y)}}
 \ee
is nothing but a classical action function of the original system [not to confuse
with the constant $I$ proportional to the action on a closed trajectory of the reduced system (\ref{caract})].
$S(x,D)$ satisfies a system of generalized Hamilton-Jacobi equations
 \be
 \label{HamJac}
\ \frac {\partial S }{ \partial D}  \frac {\partial S }{ \partial x} - DV'(x)   = E  \ , \nonumber \\
 \frac 12 \left( \frac   {\partial S}{ \partial D} \right)^2 - V(x)   = N\ .
  \ee 
For the superpotential (\ref{V24}), the second term represents the elliptic integral 
of the first kind,
 \be
E\int_0^x \frac {dy}{\sqrt{2N - \omega^2 x^2 - \frac {\lambda x^4}2}} = 
\frac {E x_0}{\sqrt{2N}} F\left(\arcsin\left(\frac x {x_0} \right), -\frac {k^2}{1-k^2} 
\right) \ .
 \ee
with $x_0$ and $k$ given above.
It is convenient to express it into inverse elliptic cosine function {\sl arccn}$(u,k)$. 
Substracting an irrelevant constant, we may 
rewrite Eq.(\ref{solHamJac}) as
 \be
\label{viaacrcn}
S(x,D) = D\sqrt{2N  - \omega^2 x^2 - \frac {\lambda x^4}2} - 
\frac {E x_0 \sqrt{1-k^2} }{\sqrt{2N}} {\rm \sl arccn}\left( \frac x{x_0} ,  k \right)\ .
 \ee
For $E,N$ satisfying the quantization condition (\ref{kvanty}), the wave function (\ref{PsiEN}) is single-valued.  

 We see that the exact solution (\ref{PsiEN})  differs from the semiclassical wave function $e^{iS}$ by the extra factor
$1/\sqrt{2N  - \omega^2 x^2 - \lambda x^4/2}$. For large enough $x$ and nonzero $N$, the function falls down 
exponentially
(we have to choose the sign of the square root in accordance with the sign of $D$). For intermediate $x$, the function
oscillates in $D$ and behaves as a plane wave continuum spectrum solution.
 When $V(x) = N$, the wave function involves
a singularity, with the normalization integral diverging logarithmically at this point. 

Two natural questions are in order now. 
 \begin{enumerate}
 \item Is this  singularity at finite  value of $x$ dangerous ? 
\item How come  the non-normalizable wave functions (\ref{PsiEN}) describe also the zero energy states ?
 The point $E=0$ is isolated
and one expects that the eigenfunctions with zero energy belong to $L^2$.
 \end{enumerate}
 
Let us answer first the second question. There are infinitely many states of zero energy. In the action-angle variables, any function
$g(I) \to \tilde g(N)$ not depending on $\varphi$ is an eigenfunction of (\ref{HamIphi}) with zero eigenvalue. In original variables, 
 this gives the function 
 \be
\label{psivac}
 \Psi_0(x,D) = (2\pi)^{-1/2} \int_{-\infty}^\infty \tilde g\left( \frac {P^2}2 - V(x) \right) {\rm e}^{iPD}dP\ .
  \ee
The solution (\ref{PsiEN}) is obtained, if substituting in Eq.(\ref{psivac})
$\tilde g(N) = \delta(N - N_0)$. But we may also choose the basis  $\tilde g_k(N) = N^k {\rm e}^{-N}$, $k = 0,1,\ldots$ (its orthogonalization 
gives the Laguerre polynomials) giving the normalized zero-energy solutions without any singularity.
Any smooth function can be expanded into this basis. 
 The distribution $\delta(N - N_0)$ can, of course, be represented as a limit 
of a sequence of smooth functions.

The existence of the normalized zero energy states together with continuum states 
could somewhat remind the maximal supersymmetric Yang-Mills quantum 
mechanics \cite{maxsusy}. There are two differences: {\it (i)} The latter is a conventional supersymmetric system and 
the zero-energy states have the meaning  of the vacuum ground
states; {\it (ii)} In our case for $\omega \neq 0$, the zero energy state is separated by a gap from continuum. For the maximal
SYM quantum mechanics, there is no gap.

The inverse square root singularity of the continuum spectrum functions  $\Psi_{E \neq 0}$ has the same nature as
the divergence of their normalization integral at large $D$. It is benign and physically admissible. Indeed, the physical
requirement for the systems with continuum spectrum is the possibility to define for any test function 
$\Psi(x,D) \in \, L^2$ the probability distribution $p(E)$, with  $p(E)dE$ giving the probability to find
 the energy of the system in the interval $[E,E+dE]$, such that the total probability integrated and/or summed over the whole
energy range is unity. This is especially clear in the action-angle representation. The requirement is that, 
  for every bounded  function $f$ and every
   test state $\Psi \in \, L^2$, the matrix element 
   \be\label{spec2}
   \langle\Psi\vert f(H_B)\vert\Psi\rangle &=& 
\sum_{n\in\Z}\int_0^{+\infty}\vert\Psi_n(I)\vert^2f(n\sigma(I))dI \nonumber\\
    &=&f(0)\int_0^{+\infty}\vert\Psi_0(I)\vert^2dI + 
\sum_{n\in\Z, n\neq 0}\int_0^{+\infty}\vert\Psi_n(I)\vert^2f(n\sigma(I))dI.
      \ee
is well defined (we have written  the contribution of the isolated spectral point $E=0$ as a distinct term). 
Now, $\Psi_n(I)$ are the Fourrier components of the test function $\Psi(I,\varphi)$. In original variables, their role
is played by the integrals
 \be
\label{genFour}
 \int \, \Psi(x,D) \Psi_{EN}(x,D)\, dx dD\ ,
 \ee  
These integrals converge (though the normalization integrals for $\Psi_{EN}(x,D)$ do not) 
and the weak singularity $\propto 1/\sqrt{x-x_0}$ does not hinder this convergence.

\subsection{Including fermions}

Once the bosonic problem is resolved, it is not difficult to obtain the solution of the full problem
(\ref{H}). To be more precise, the Hamiltonian (\ref{H}) as it is may appear problematic. For example, it
does not look Hermitian --- not invariant  under the conjugation 
$\psi \leftrightarrow \bar \psi, \chi \leftrightarrow \bar\chi$
    \footnote{Note, however, that  it is invariant
  with respect to the involution
 \be
\label{invol}
\psi \to \chi, \ \ \chi \to -\psi, \ \  \bar\psi \to \bar \chi, \ \ \bar \chi  \to -\bar\psi, \ \
 \ee
supplemented by the usual complex conjugation of the bosonic variables. }.   

%Note first of all that the time--dependent Schr\"odinger equation can be easily resolved by the same method as in the bosonic case.
%We introduce $\eta = \bar\chi, \bar\eta = \chi$ and use the variables $(x,P, \psi, \eta)$. The Schr\"odinger equation takes the form 
 % \be
%\label{Schferm}
 % i\frac{\partial \Psi}{\partial t} + iP\frac{\partial \Psi}{\partial x} + 
%iV^\prime(x)\frac{\partial  \Psi}{\partial P} + \eta \frac {\partial \Psi}{\partial \psi} - \psi V''(x) \frac {\partial \Psi}{\partial \eta} = 0
%\,.
 % \ee
%Again, this is a homogeneous linear first order differential equation and its solution can be written in analogy with (\ref{Sch2})
 %\be\label{fermflow}
%\  \Psi_t(x,P;\psi, \eta) = \  \Psi_0(\Gamma^{-t}(x,P; \psi, \eta))\, ,
%\ee
%where $\Gamma^t$ is now the flow of the characteristic system involving besides (\ref{caract}) also the equations for the fermion variables,
% \be
%\dot \psi &=& -i\eta\, , \nonumber \\
%\dot \eta &=& iV''(x) \psi \ .
% \ee
%This proves that the evolution operator is unitary and the Hamiltonian is Hermitian in $L^2(\R^2) 
%\otimes \Lambda(\C^2)$.

Still, the spectrum of the Hamiltonian can be found by supersymmetry considerations.
 The states are classified by the value of the fermionic charge $F$, which can take values $-1,0,1$. 
The wave functions of the states in the sectors $F = -1$ and $F = 1$ involve the factor $\chi$ and $\psi$, 
correspondingly. The fermion
part of the Hamiltonian does not act on such states and the solutions to the stationary Schr\"odinger 
equation in these sectors can be 
immediately written,
 \be
\label{Fpm1}
 \Psi^{F = -1}(x,D; \psi, \chi) &=& \chi \Psi_B(x,D)\, , \nonumber \\
  \Psi^{F = 1}(x,D; \psi, \chi) &=& \psi \Psi_B(x,D)
\ee
with $\Psi_B$ written above in Eq.(\ref{PsiEN}).
The eigenstates of the same energy in the sector $F=0$ can be obtained from the states (\ref{Fpm1}) 
by the action of the 
supercharges $Q,\bar Q, T, \bar T$,
 which commute with the Hamiltonian. There are two such states,
\be
 \label{F01}
 \Psi_1^{F = 0}(x,D; \psi, \chi)  = \frac{T - Q}2 \Psi^{F= -1}  = \frac{\bar T  + \bar Q}2 \Psi^{F= 1} \ =
\nonumber \\  
\left[\sqrt{2[N+V(x)]} -iV'(x) \psi\chi  \right]\Psi_B(x,D)\, .
   \ee
and 
 \be
 \label{F02}
 \Psi^{F = 0}_2(x,D; \psi, \chi)  = \frac{Q+T}2 \Psi^{F= -1} = \frac{\bar Q  - \bar T}2 \Psi^{F= 1} \ =\ 
i\left[D - \psi\chi \frac \partial{\partial x}  \right]\Psi_B(x,D)\, .
   \ee
The action of $Q,T$ on $\Psi^{F= 1}$ and the action of $\bar Q, \bar T$ on $\Psi^{F= -1}$ give zero. 

And here we meet a certain difficulty. As far as the states (\ref{F01}) are concerned, everything is fine,
these states behave in the same way as $\Psi_B$, involving a benign integrable singularity 
$\propto 1/\sqrt{x-x_0}$. The generalized Fourrier integrals (\ref{genFour}) are finite, and 
the probability for a 
test function $\Psi$ from ${\cal L}_2$ to have an energy within a given interval $(E, E+dE)$ is well defined.
However, the functions (\ref{F02}) are more troublesome. They involve a singularity $\propto 1/(x-x_0)^{3/2}$,
which is not integrable in the usual sense. 
                        
We have two options now:
\begin{enumerate}
\item To declare the state (\ref{F02}) inadmissible by that reason. In that case, the spectrum would lose its 
supersymmetric form and would involve not quartets, but {\it triplets} of states of given energy, as 
illustrated in Fig.\ref{triplet}.
 \begin{figure}[h]
   \begin{center}
 \includegraphics[width=4.0in]{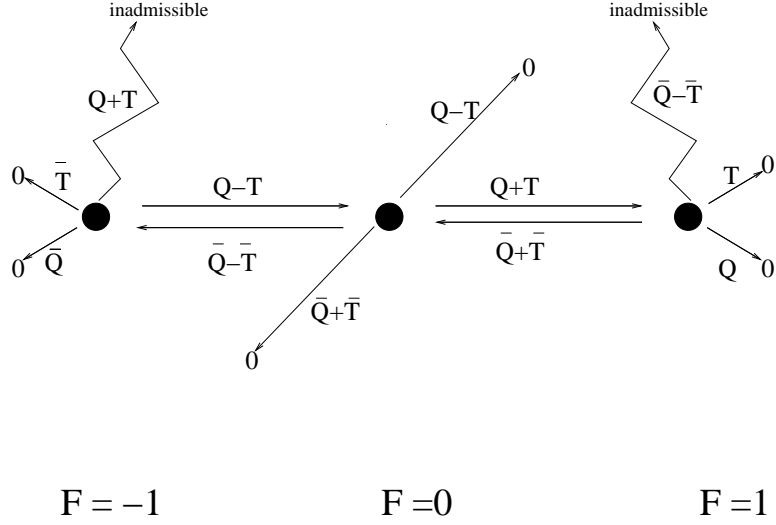}
      %\vspace{-1cm}
  \end{center}
  \caption{The triplet of  ``benign'' states under the action of  supercharges}
  \label{triplet}
  \end{figure}

%\newpage 
\item The second option is to {\it include} the states (\ref{F02}) into  some extended  Hilbert space and to {\it define} 
the integrals (\ref{genFour}) with a kind of principal value prescription. Roughly speaking, we represent 
 \be
\label{x32} 
I = \int \Psi(x) \frac {dx} { x^{3/2}} \ \ \ \ {\rm as} \ \ \ \ 
-2\int dx \, \Psi(x)  \frac {\partial }{\partial x} \frac {1} {\sqrt{x}} 
 \ee
(recall that the structure $\sim 1/x^{3/2}$ appeared in the first place when applying the differential operator
to $\Psi_B$). We can integrate then by parts and write
$$I = 2 \int \Psi'(x)  \frac {dx} {\sqrt{x}}$$
 with an integrable singularity. For mathematical precision, we can
regularize 
$$ \frac 1{\sqrt{x}} \ \to \ \left.   \frac 1{\sqrt{x}} \right|_\epsilon \ =\ 
\frac 1{\sqrt{\pi} \epsilon}  \int dy \, 
\frac 1{\sqrt{x-y}} \,  \exp\{-y^2/\epsilon^2\}\, ,$$
do the integral in the R.H.S. of Eq.(\ref{x32}) at finite $\epsilon$ and take 
the limit $\epsilon \to 0$  afterwards. 
\end{enumerate} 

The second option seems to us more reasonable. Once there is a consistent 
way to ascribe meaning to the states and the associated generalized Fourrier integrals (\ref{genFour}) such that 
one of the important symmetries of the problem, the supersymmetry, is kept intact, it is preferrable 
to use it.   In addition, the system of eigenfunctions involving only the states (\ref{Fpm1}) and (\ref{F01}), but
not (\ref{F02}) does not seem to be complete 
\footnote{We cannot formulate it as a rigourous mathematical statement, because we are dealing with continuum spectrum
here,  
and the states (\ref{Fpm1},\ref{F01},\ref{F02}) do not represent a conventional basis in a conventional Hilbert space. The usual
way to treat continuous spectrum systems is to put them in a finite box to make the motion finite and the spectrum
discrete. The problem is that a na\"ive box regularization breaks supersymmetry. More study of this question is required.}      

 As was mentioned before the Hamiltonian (\ref{H}) is  not Hermitian. Still, its spectrum is  real. 
That means that this Hamiltonian belongs to the class of so-called quasi-Hermitian or crypto-Hermitian
Hamiltonians \cite{quasi,Bender,crypto}.

%%%%%%%%%%%%%%%%%%%%%%%%
%%%%%%%%%%%%%%%%%%%%%%%

\section{Models in the neighborhood.}
The Lagrangian (\ref{supact}) is the simplest nontrivial supersymmetric Lagrangian with higher derivatives. But there are many other such theories.
In this section, we discuss two different natural modifications of (\ref{supact}) 

\subsection{Mixed theory.}
One obvious thing to do is add to (\ref{supact}) the standard kinetic term multiplied by some coefficient $\gamma$ and write
\be
\label{mixed}
L \ = \ \int  d\bar \theta d\theta \left[ \frac i2 ( \bar {\cal D} X) \frac d{dt} ({\cal D} X) + \frac \gamma 2   \bar {\cal D} X 
{\cal D} X  + V(X) \right]\ .
 \ee
The component expression for the Lagrangian  is
 \be
\label{Lmixed}
L \ =\ \dot x \dot D + D V'(x) + V''(x) \chi \psi + \dot \chi \dot \psi + %\nonumber \\
\gamma \left[ \frac {\dot x^2 +  D^2}2 +
 \frac i2 (\dot \psi \chi - \psi \dot \chi ) \right] \ .
 \ee
The canonical Hamiltonian is convenient to express as 
\be
\label{H0F}
H = H_0 - \frac \gamma 2 F\ ,
 \ee
where $F$ is the operator of fermion charge and 
 \be
\label{H0}
H_0 = pP -DV'(x) - \frac \gamma 2 (D^2 + P^2) + \bar \psi \bar \chi + 
\left[ \frac {\gamma^2}{4}  - V''(x) \right] \chi \psi  \ .
 \ee
One can find out also the N\"other supercharges $Q, \bar Q$. Being expressed via canonical momenta, they are
 \be
\label{Qgam}
Q &=& \psi[p + iV'(x)] - \left(\bar \chi + \frac \gamma 2 \psi \right)(P-iD) \, , \nonumber \\
\bar Q &=& - \chi[p -iV'(x)] + \left( \bar \psi + \frac \gamma 2 \chi \right) (P + iD) \, .
 \ee
Further, one can {\it guess} the existence of the following generalization for the second pair of the supercharges $T, \bar T$,
 \be
\label{Tgam}
T &=& \psi[p - iV'(x)] + \left(\bar \chi - \frac \gamma 2 \psi \right)(P+iD) \, , \nonumber \\
\bar T &=&  \chi[p + iV'(x)] + \left( \bar \psi - \frac \gamma 2 \chi \right) (P - iD) \, .
 \ee
Introducing also the operators $F_+ = \bar \chi \psi$ and $F_- = \bar \psi \chi $, 
one can observe that the superalgebra of the set of the operators
$H_0, F, F_+, F_-;\ Q, \bar Q, T, \bar T$ is closed. The nonvanishing (anti)commutators are
 \be
\label{alggam}
 \ [F_\pm, F] = \mp 2F_\pm,\ \ [F_+, F_-] = F\ , \nonumber \\
\ [Q,H_0] = -\frac {\gamma} 2 Q, \  [\bar Q,H_0] = \frac \gamma 2 \bar Q, \ [T,H_0] = \frac 
\gamma 2 T, \ [\bar T, H_0] = -\frac \gamma 2 \bar T
\, , \nonumber \\
\ [Q,F] = -Q, \ \ [\bar Q, F] = \bar Q,\ \ [T,F] = T, \ \ [\bar T, F] = -\bar T \, ,  \nonumber \\
\ [Q, F_-] = \bar T, \ \ [\bar Q, F_+] = -T,\ \ [T, F_-] = - \bar Q, \ \ [\bar T, F_+] = Q \, ,  
\nonumber \\
\ \{Q, \bar Q\} = 2H_0 - \gamma F,\ \ \{T, \bar T\} = 2H_0 + \gamma F,\ \ 
 \{Q,T\} = 2\gamma F_+,\ \ \{\bar Q, \bar T\} = 2\gamma F_-\ .
 \ee
One can make here a few remarks. 
\begin{itemize}
 \item  The operators $F, F_+, F_-$ form the $sl(2)$ subalgebra. 
 One could have introduced
the operators $F_\pm$ also in the case $\gamma = 0$, but that was not necessary for closing the algebra. 
When $\gamma \neq 0$, it
is. Actually, Eq.(\ref{alggam}) represents a well known simple superalgebra 
$sl(1,2) \equiv osp(2,2)$ \cite{Kac}.  
 \item  The algebra (\ref{alggam}) involves two conventional 
${\cal N}=1$ subalgebras. They are realized by
the subsets ($ H_0 - (\gamma/2) F; Q, \bar Q$) and ($ H_0 + (\gamma/2) F; T, \bar T$). Recall, however,
 that the operators $Q, \bar Q$ and 
$T, \bar T$ are not Hermitially conjugate to each other, which allows for the presence of the negative 
energies in the spectrum.
 \item  The algebra (\ref{alggam}) is a close relative of
 the unconventional so called {\it weak} supersymmetric algebra introduced in  Ref.\cite{weak}. 
One can show that the latter is a semidirect sum 
of the algebra (\ref{alggam}) with the Abelian 1-dimensional algebra $(Y)$. 
 \item  When $\gamma \to 0$ , the subalgebra of  (\ref{alggam}) 
involving only the operators $H_0 \equiv H, F;\ Q,T, \bar Q, \bar T$ coincides with the subalgebra 
of (\ref{supalgebra}) involving the same operators. The system with $\gamma = 0$ involves an additional 
integral of motion $N$, but,  when $\gamma \neq 0$, the Lagrangian is not invariant with respect to the transformations
(\ref{symact}) anymore and 
there seems to be no such integral.
 \end{itemize}

What is the dynamic of the mixed  system ? Consider first the classical bosonic dynamics. The Hamilton equations of motion  are now
 \be
\label{eqmotgam}
\dot p &=& D V''(x)\,, \nonumber \\
\dot P &=& V'(x) + \gamma D \,, \nonumber \\
\dot x &=& P \,, \nonumber \\
\dot D &=& p - \gamma P\, .
 \ee

\begin{figure}[h]
   \begin{center}
 \includegraphics[width=9.0in]{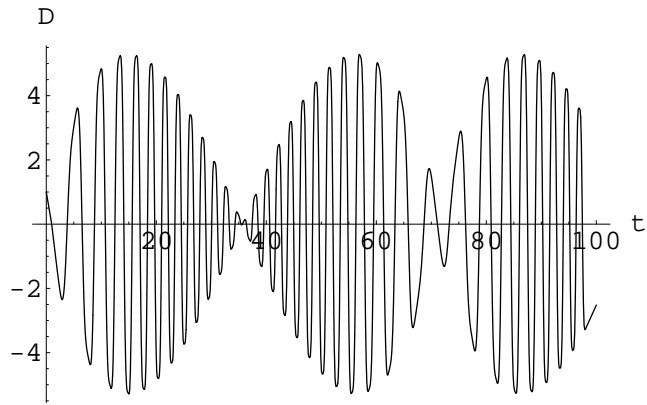}
        \vspace{-22cm}
    \end{center}
\caption{The function $D(t)$ for a deformed system ( $\omega =0, \lambda =1, \gamma =.1$).}
\label{bienie}
\end{figure}

The absence of the extra integral of motion $N$ seems to make
 the system  not integrable.    
\footnote{It is  difficult to prove the {\it absence} of something. One can always say that the extra integral actually exists, but we simply
have not found it. We performed, however, a numerical study which {\it suggests} that the system is not integrable. In particular, when 
$\gamma \neq 0$, the parametric plot of the solution in the plane $(x,P)$ does not represent a closed curve as it does for $\gamma = 0$, but
densely covers a certain region in the phase space.} 
That means that analytic solutions do not exist, but it is possible to study the solutions
numerically. Remarkably, it turns out that the trajectories are in this case in some sense more benign that for undeformed system. 
When $\gamma = 0$, the function $x(t) = x_0 {\rm cn}[\Omega t, k]$ varies within a finite region, but the amplitude of the oscillations for 
$D(t)$ grows linearly in time (see Eqs.(\ref{D2t},  \ref{solD}) and Fig.1). 
For nonzero $\gamma$, it does not and the motion is {\it finite}. When $\gamma$ is small, 
the amplitude pulsates as is shown in Fig.\ref{bienie}. The larger is $\gamma$, the less is the amplitude and the period of these 
pulsations. When $\gamma$ is large, the ``carrying frequency'' and amplitude    fluctuate in an irregular way.    
 
\vspace{-3cm} 

What can one say about the structure of the spectrum ? The finiteness of motion suggests that the spectrum might be discrete. Let us show 
that it  {\it is} not  discrete in the precise mathematical meaning of this word. A discrete spectrum involves only
{\it isolated} eigenvalues \cite{Simon}.
We will prove, however, that the system under consideration   
involves an infinite number of eigenvalues in any finite energy interval.
 
Let us consider first the sectors $F = \pm1$ where the problem is equivalent to 
 a purely bosonic problem with the Hamiltonian
 \be
\label{HBgam}
H_B \ =\  pP -DV'(x) - \frac \gamma 2 (D^2 + P^2)\ .
 \ee
%%%%%%%%%%%%%%
%%%%%%%%%%%%%%%
$ H_B$ is unitary equivalent, up  to a partial Fourier transform, to 
\be
 K_B = \frac{1}{i}\left(P\partial_x+V^\prime(x)\partial_P\right) -\frac \gamma 2 (P^2  - \partial_P^2),
\ee
where $\partial_P = \frac{\partial}{\partial P}$.\\
  We can see that  $K_B$  has the same algebraic structure as the Fokker-Planck operator 
\footnote{We thank B. Helffer for this remark.}  but there is a big difference:
 the Fokker-Planck operator $ H_{FP}$  is 
 $$
 H_{FP}^{V'} = y\partial_x+V^\prime(x)\partial_y -\frac \gamma 2 (y^2-\partial_y^2)
 $$
 so $H_{FP}$ is not Hermitean but the Hermitean   part of $H_{FP}$ is negative ($\gamma > 0$) or positive ($\gamma <0$) definite.  
Here  $ K_B$ is Hermitean and is unbounded both from above and from below,
 and we have lost the ``hypoelliptic character"  of the Fokker-Planck operator (see \cite{HN} for more details).

  Nevertheless, for harmonic potentials $V(x) = -\frac{\omega^2x^2}{2}$,  we can compute  the spectrum of 
 $H_B$ explicitely. To  diagonalize  the (classical) Hamiltonian (\ref{HBgam}), we introduce first canonically conjugated
holomorphic variables
 \be
 a = \ \frac{p- i\omega^2 x}{\omega \sqrt 2}, \ \  a^* = \ \frac{p+ i\omega^2 x}{\omega \sqrt 2} \ ; \ \ \ \ \ \ \ 
 b = \ \frac{ P -iD}{\sqrt 2}, \ \  b^* = \frac{ P + iD}{\sqrt 2}\  ,
  \ee
 such that the Hamiltonian acquires the form 
  $$
   H_B = \omega (ba^* + ab^*) - 
\gamma bb^* \ .
  $$
Then we rotate 
\be
\label{rotation}
\left( \begin{array}{c} c_1 \\ c_2 \end{array} \right) \ =\ 
\left( \begin{array}{cc} \cos \phi & \sin \phi \\ -\sin \phi & \cos \phi \end{array} \right) 
 \left( \begin{array}{c} a \\ b \end{array} \right)
  \ee
with $\tan (2 \phi) = 2\omega/\gamma \equiv \tau$. The Hamiltonian is reduced to 
\be
\label{HBdiag}
H_B \ =\ \omega_1 c_1 c_1^* - \omega_2 c_2 c_2^* \ ,
 \ee
where
$$ \omega_{1,2} = \frac \gamma 2 \left( \sqrt{1 + \tau^2} \mp 1 \right) \ . $$
 The spectrum
  \be
\label{spect}
  \omega_{jk} \ =\ \left(\omega_1+\frac{1}{2} \right)j -  \left(\omega_2+\frac{1}{2} \right)k\  + \ const, \ \ \ \ \ \ \  j, k\in \N ,
  \ee
 of the quantum counterpart of this Hamiltonian 
is pure point ( a  constant shift in (\ref{spect}) is inserted for generity. It  takes into account  the ordering ambiguity).     
  Moreover, this spectrum is dense  if $\omega_1 /\omega_2$ is not rational, which is true for 
generic $\tau$.

It is worth noting that the system (\ref{HBgam}) with quadratic $V(x)$ is  equivalent to {\it  Pais-Uhlenbeck oscillator} 
--- a higher-derivative dynamic system described by the Lagrangian \cite{PU}
 \be
\label{PUosc}
L \ =\ \frac 12 \left( \ddot {q}^2 - (\omega_1^2 + \omega_2^2) \dot{q}^2 + \omega_1^2 \omega_2^2 q \right) 
 \ee
The Hamiltonian can be found using Ostrogradsky formalism \cite{Ostr}. If quantizing it and evaluating the spectrum, we obtain
exactly  the same result as in Eq.(\ref{spect}) \cite{DM,dno}.

 {\bf Remark}:
  For more general potential $V$,  it is not immediately   clear that $ H_B$ is essentially self-adjoint.
  This can be proved if $V''(x)$ is uniformly bound  on $\R$. But we do not know how to obtain rigourous results
about  the spectrum of $H_B$ for non quadratic potentials. 
If  the coupling $\gamma$ is  imaginary, we are in  the Fokker-Planck  situation and,  if some confining  conditions on the 
potential $V$ are satisfied,   it is known then that the spectrum of $ K_B$ is pure point with bounded from below eigenvalues.
  Moreover,  in this case  the resolvent of $ K_B$ is compact \cite{HN}.

%%%%%%%%%%%%%
%%%%%%%%%%%%%
For real positive $\gamma$ and more general potentials $V$ a semi-heuristic reasoning revealing the nature of the spectrum
can be presented. 
Consider the quantity $Z[f] = {\rm Tr}\{f(H_B) \}$, where $f(u)$ is any positive definite function dying at $u = \pm \infty$ fast enough. 
For example,
one can take $f(u) = \exp\{-u^2/\sigma^2\}$. In semiclassical approximation, we can evaluate it as
 \be
 \label{Zsemicl}
Z[f] \approx \int \frac {dx dp dD dP}{(2\pi)^2} \ f[H_B^{\rm cl}(x,D,p,P) ]\ ,
 \ee
where $H_B^{\rm cl}$ is the Weyl symbol of the quantum Hamiltonian [in our case, it is given directly by Eq.(\ref{HBgam})].
The corrections to this formula [their existence can be understood by noting that the Weyl symbol of $f(\hat H_B)$ does not coincide
with $f(H_B^{\rm cl})$] can also be evaluated \cite{HelRob}. When the function $f(u)$ is smooth enough 
(for $f(u) = \exp\{-u^2/\sigma^2\}$
the condition is  $\sigma \gg 1$), the corrections are small. Doing in Eq.(\ref{Zsemicl}) the integral over $dDdP$, we obtain
 \be
\label{dxdp}
Z[f] \approx \int \frac {dxdp}{2\pi\gamma} \ g \left( \frac {p^2}{2\gamma} + \frac {[V'(x)]^2}{2\gamma} \right)
 \ee
with
$$ g(u) = \int_{-\infty}^u f(w) dw \ .$$
When $x$ and/or $p$ and hence $u$ are large, the integrand in (\ref{dxdp}) is a constant, and the integral diverges. On the other hand, assuming
that the spectrum is pure point, one may write
 \be
\label{Zsum}
Z[f] \ =\ \sum_n f(E_n)
 \ee
The infinite value of this sum for any function $f$ including the functions that die at infinity very fast 
means the presence of an infinite number of states in a finite energy range. That means that the spectrum should have
accumulation points so that the eigenvalues of the Hamiltonian are not well separated from each other. It is conceivable
that in our case {\it each} point of the spectrum is an accumulation point and the spectrum represents a countable
subset of $\R$ that is dense everywhere. This is what happens for simpler models with the Hamiltonian like
 \be
\tilde{H} = \frac {p^2}{2\gamma} + \frac {[V'(x)]^2}{2\gamma} - \frac \gamma 2 (P^2 + D^2)
 \ee
(the semiclassical value of $Z[f]$ for the Hamiltonian $\tilde{H}$ is the same as for $H_B$).

Another possibility is that the spectrum of $H_B$ is truly continuous with not normalizable wave functions. Based on the mentioned above 
fact that the classical motion of our system is finite, we find this option less probable. But only a future study will allow one
to obtain a definite answer to this question.

Let us briefly discuss the dynamics of the full supersymmetric system. As was explained above, it involves two pairs of complex supercharges. However,
the supercharges $T, \bar T$ do
{\it not} commute with the Hamiltonian $H \equiv H_- = H_0 - (\gamma/2)F$, but only with the operator $H_+ = H_0 + (\gamma/2) F$.
 On the other hand, $Q$ and $\bar Q$ do not commute with $H_+$. 
The fact that the maximal conventional supersymmetry subalgebra involves in this case
only one pair of complex supercharges implies that the 
  spectrum of $H$ consists of degenerate doublets rather than quartets (as was the case for $\gamma=0$). 
\footnote{The commutation relation $[T,H] = \gamma T$ guarantees that, 
if $\Psi$ is the eigenstate of $H$, \ $T\Psi$ is also
an eigenstate, but with a different eigenvalue. Thus, only the supercharges $Q, \bar Q$ are effective as far as degeneracy is concerned. } 
The same concerns $H_+$. 

 The doublet structure of the spectrum is a  feature which distinguishes
  the system under consideration from 
 the system considered in \cite{weak}. 
The algebra of the latter was similar to (\ref{alggam}), but involved an extra bosonic charge $Y$. That allowed
for the existence of  an operator  that commutes 
with all supercharges. The spectrum of this operator (it is natural to call it Hamiltonian) 
beyond the ground state and the first excited state is  4-fold degenerate.

\subsection{More  derivatives.}

 As a final example, consider  a somewhat more complicated action
 \be
\label{supact2}
S \ = \ \int dt d\bar \theta d\theta \left[ \frac 12 ( \bar {\cal D} \dot X) \, ({\cal D} \dot X) +
V(X) \right]\ .
 \ee
 The corresponding component Lagrangian is
 \be
\label{L2dot}
 L \ =\ \frac 12 \left[ \ddot x^2 + \dot D^2 \right] + i \ddot{\bar \psi} \dot \psi + DV'(x) +
V''(x) \bar \psi \psi \ .
  \ee
The bosonic equations of motion are
 \be
 \label{eqmotdot2}
  x^{(4)} + D V''(x) &=& 0\,, \nonumber \\
 \ddot D - V'(x) &=& 0\ .
 \ee
In the simplest quadratic case
 \be
\label{om3}
V(X) \ =\ \frac {\omega^3 X^2}2\ ,
 \ee
the equations (\ref{eqmotdot2}) are linear and can  readily be solved. Their characteristic 
eigenvalues are 
 \be
\lambda_{1,2} = \pm i\omega,\ \ \ \ \ 
\lambda_{3,4,5,6} = \omega\left( \pm \frac {\sqrt{3}}2 \pm \frac i2 \right)\ .
 \ee
We see that, besides oscillating solutions, there are also solutions with exponentially growing amplitude.
Strictly speaking, the Hamiltonian is still Hermitian in a certain sense   due to the fact that there is no collapse:
it takes an infinite time to reach infinity and a unitary evolution operator can be defined at all times.
However, Hermiticity is lost
\footnote{modulo a possible remedy in the spirit of \cite{Bender}, see the discussion below.} 
 as soon as one switches on interactions. We solved numerically the equations of
motion (\ref{eqmotdot2}) for $V(x) \propto \pm x^3$ and  $V(x) \propto \pm x^4$ and found out that the solutions
collapse reaching a singularity at finite time.

\section{Discussion}

Probably, the main lesson to be learned from the analysis of different higher-derivative quantum mechanical
models in this paper is that the ghosts (negative energy states and the Hamiltonians without bottom) do not {\it always} lead
to violation of unitarity, but one should worry about it only in the case when the collapsing classical trajectories exist.
The analysis performed in Refs.\cite{benign,5d} displays that sometimes even in this case the quantum problem 
 is (or can be, if defining the Hilbert space with a care) well defined, but for the model (\ref{supact}) where there is 
no collapse, quantum evolution is unitary in spite of the absence of the ground state. 

In addition to this, we found a bunch of rather unusual phenomena.
\footnote{They are unusual for conventional systems with positive definite kinetic term, but maybe not so unusual for the system
involving ghosts. Unfortunately, the latter were never seriously studied before.}
 \begin{enumerate}
 \item The classical trajectories of the system (\ref{supact}) do not collapse, but exhibit oscillatory behavior
with linearly rising amplitude.  For the modified model
(\ref{mixed}), the amplitude does not grow and the trajectories  are {\it finite}.
 \item The model (\ref{supact}) is exactly soluble due to the presence of an extra integral of motion. For the quartic potential, 
the solutions of the classical and quantum problems are expressed analytically via elliptic functions.  
 \item Besides N\"other supercharges, the systems (\ref{supact}) and (\ref{mixed}) involve an extra pair of supercharges. N\"other
supercharges, the additional supercharges, the Hamiltonian and certain extra operators form  a modified supersymmetry algebra. 
For different systems, these modified superalgebras are also different. 
 \item The system (\ref{supact}) has continuous spectrum. The spectrum of the model (\ref{mixed}) is probably not continuous, but
involves only normalizable discrete spectrum states, with a countable set of eigenvalues densely covering $\R$. This conjecture
needs to be confirmed.  
  \item The spectrum of the Hamiltonian (\ref{H}) was found to be real even though this Hamiltonian is na\"ively not Hermitian
in the fermion sector. That means that it belongs to the class of so called quasi-Hermitian, alias crypto-Hermitian Hamiltonians
having attracted recently a considerable interest \cite{quasi,Bender,crypto}.
\footnote{In supersymmetric context, crypto-Hermitian systems were first discussed in \cite{Mostaf}. It was found recently that
the Hamiltonians describing so called nonanticommutative supersymmetric theories \cite{Seiberg} are crypto-Hermitian \cite{ASIS,cryptosusy}.
 Non-Hermitian supersymmetric $\sigma$ models, having certain kinship to 
 our models, were considered  recently \cite{Nekrasov}.    } 
A salient feature of  such systems is the possibility to define a modified
norm in Hilbert space such that the Hamiltonian is manifestly Hermitian with respect to this norm  \cite{Ali}.
 It would be interesting to find out whether 
such modified norm exists also in our case. 

 \end{enumerate}

The central question posed in Refs.\cite{benign,TOE},  whether benign  higher-dimensional higher-derivative supersymmetric 
field theories exist, is still left unresolved. The best currently known 
candidate for this role is  superconformal at the classical
level renormalizable gauge theory in six dimensions constructed in Ref.\cite{ISZ}.
 Unfortunately, it is not free of difficulties and it is not clear at the moment whether they can be resolved or not.
The simplest such  model constructed in \cite{ISZ} is probably not viable because of chiral anomaly. 
However, the anomaly is cancelled in a theory involving besides gauge supermultiplet also 
a matter hypermultiplet \cite{anom}. On the other hand,  the  latter scale-invariant theory 
is not fully  conformal even at the classical level and
has an infinite number of propagating fields \cite{hyper}. 

In addition,
in contrast to the models discussed in the present  paper, the models \cite{ISZ,hyper} {\it do} involve collapsing classical 
trajectories reaching infinity in finite time. This is due to the presence of the cubic term $\sim D^3$ 
in the Lagrangian. 
(Fields $D$ are the highest components of the vector ${\cal N} = 1$ $6D$ supermultiplet of canonical dimension 2. 
They are auxiliary for the standard quadratic in derivatives theory, but become dynamical
when extra derivatives are added.) The toy models analyzed  above provide no care for such a theory.
The latter still {\it might} 
 exist in some sense. Indeed, if the theory with complex cubic potential  $V(x) \propto ix^3$ involves 
a benign crypto-Hermitian Hamiltonian \cite{Bender}, the theory with real cubic potential $\propto x^3$ may  also
acquire sense if considering it on the complex $x$-plane 
\footnote{The idea to get rid of ghosts by rotation in complex plane was put forward 
in \cite{BM}. In was based on the simple observation that the spectrum of a standard harmonic oscillator changes
sign under rotation $x \to ix$.  Consider then the Pais-Uhlenbeck oscillator (\ref{PUosc}) presented in the form
(\ref{HBdiag}) and perform such rotation for the {\it second} subsystem. We will obtain instead of (\ref{spect})
a positive definite ghost-free 
spectrum $\omega_{jk} =  \omega_1 j + \omega_2 k $ + const. In our opinion, in the case of the free 
PU oscillator, there is no need to do such transformation. As explained above, the system with the spectrum 
(\ref{spect}) makes perfect sense, the ghosts are ``benign''. Complex rotations may prove useful, however, 
for interacting systems.}

 It is not 
inconceivable that the higher derivative theory of Ref.\cite{ISZ} can also be treated along these lines. 
On the other hand, if we are interested not just in quantum mechanical
systems, but in field theory, we would like to have not only Hermitian Hamiltonian, but also unitary $S$-matrix.
A discouraging news \cite{cryptosusy} is that, for the field theories involving the cubic term in the potential,
the unitarity of $S$-matrix  is impossible to preserve, if choosing the asymptotic states in a conventional way.
It is not clear, however, whether we {\it have} 
 to require the existence of conventional asymptotic
states and the conventional unitary $S$ - matrix for the fundamental theory in the higher-dimensional bulk. Maybe,
it would suffice to have a unitary finite time evolution operator with a certain (complicated) choice for the Hilbert
space metric ?   
 
Further studies of all these questions are necessary.  

We are indebted to E. Ivanov, B. Helffer, V. Kac, M. Kroyter, N. Nekrasov,  and S. Theisen  
for illuminating discussions and correspondence.

\end{document}